\documentclass[conference]{IEEEtran}
\IEEEoverridecommandlockouts
\usepackage{cite}
\usepackage{amsmath,amssymb,amsfonts}
\usepackage{bm}
\usepackage{algorithm}
\usepackage{algorithmic}
\usepackage[T1]{fontenc}
\usepackage{graphicx}
\usepackage{textcomp}
\usepackage{xcolor}
\usepackage{newtxtext,newtxmath}
\def\BibTeX{{\rm B\kern-.05em{\sc i\kern-.025em b}\kern-.08em
    T\kern-.1667em\lower.7ex\hbox{E}\kern-.125emX}}

\newtheorem{theorem}{Theorem}
\newtheorem{lemma}{Lemma}

\begin{document}
\title{Design of OTFS Signals with Pulse Shaping and Window Function for OTFS-Based Radar
Systems
\thanks{This work was supported in part by JSPS KAKENHI Number JP23K26104 and JP23H00474.}
}
\author{\IEEEauthorblockN{Liangchen Sun}
\IEEEauthorblockA{\textit{Department of Information Science and Technology} \\
\textit{Kyushu University, Japan}\\
sun@me.inf.kyushu-u.ac.jp}
\and
\IEEEauthorblockN{Yutaka Jitsumatsu}
\IEEEauthorblockA{\textit{Deptartment of Informatics,} \\
\textit{Kyushu University, Japan}\\
jitsumatsu@inf.kyushu-u.ac.jp}
}

\maketitle
\begin{abstract}

We propose a pulse radar system that employs a generalized window function derived from the root raised cosine (RRC), which relaxes the conventional constraint that the window values are within the range [0, 1]. The proposed window allows both negative values and values exceeding 1, enabling greater flexibility in signal design. The system transmits orthogonal time frequency space (OTFS) signals intermittently, establishing a flexible input–output relationship that captures both fractional delays and Doppler shifts. By combining the generalized RRC window with a rectangular pulse, the resulting pilot signal achieves a sharp concentration in the ambiguity function over both the delay and Doppler domains. To enhance the estimation accuracy of fractional parameters, we apply frequency-domain interpolation based on the autocorrelation of the RRC window, which outperforms conventional linear interpolation by preserving the signal structure more effectively.

\end{abstract}

\begin{IEEEkeywords}

OTFS, radar, delay Doppler estimation

\end{IEEEkeywords}
 
\section{Introduction}

In recent years, significant progress has been made in the realization of wireless communication using millimeter waves, despite their high attenuation and strong directive, which make them susceptible to obstructions ~\cite{millimeter-wave2021}.  
Millimeter waves, on the other hand, have long been used in radar frequency bands. Building on this, there is growing interest in extending automotive radar systems to incorporate communication capabilities, enabling vehicle-to-vehicle (V2V), vehicle-to-infrastructure (V2I), and more broadly, Vehicle-to-Everything (V2X) communication. In this context, Joint Communication and Sensing (JCAS), which uses a single modulation signal to achieve both communication and sensing functions, has gained significant attention~\cite{jcas,jrac}. From an information-theoretic perspective, the integration of communication and sensing within the same frequency band is more efficient than the orthogonal division of the band to implement these functions independently~\cite{MariKobayashi2018, CaireOTFS2022}. In JCAS research, OTFS (Orthogonal Time-Frequency Space) signals have gradually become a hot research topic due to their robustness in time-frequency dispersive channels and strong resistance to multipath and Doppler effects. Although Frequency Modulated Continuous Wave (FMCW) remains the mainstream solution for automotive radar, OTFS is expected to demonstrate superior performance in this field due to its excellent anti-interference capability.

Although extensive research has been conducted on OTFS-based radar systems, many approaches still face challenges, particularly in fractional delay and Doppler estimation ~\cite{Gaudio2020, dehkordi2022beam, Zhang2023Radar, wu2023dft, zacharia2023fractional, zielinski2024wireless}. While some studies propose potential solutions, most adopt the rectangular window and overlook the impact of window design on system performance~\cite{dehkordi2022beam, Zhang2023Radar, wu2023dft, zacharia2023fractional}. Compared to the rectangular window, well-designed generalized window functions can significantly reduce spectral leakage, control noise characteristics, and manage computational complexity~\cite{farhang2017low}. Furthermore, given sufficient computational resources, generalized window functions can provide additional degrees of freedom to enhance performance ~\cite{raviteja2018interference}. Although recent studies have explored window function design in OTFS communication, their potential in sensing remains  unexplored~\cite{wei2021transmitter}.

\begin{figure}
    \centering
    \includegraphics[width=1\linewidth]{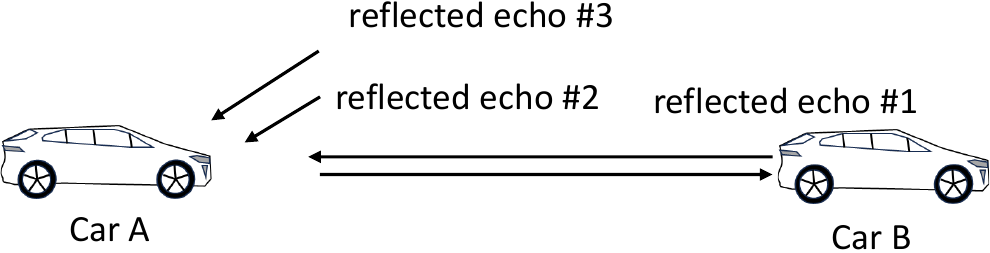}
    \caption{A vehicle with radar/radio communication system and reflected echoes}
    \label{fig:Car_and_reflected_echos}
\end{figure}

To address these limitations, this paper proposes a V2V communication system that simultaneously measures the distance and relative velocity of preceding vehicles using reflected waves, while enabling data exchange (Fig. \ref{fig:Car_and_reflected_echos}). To facilitate explanation, we assume a single-user scenario. Our contribution lies in the application of RRC (root raised cosine) window functions in OTFS signal design, which, to the best of our knowledge, is the first such attempt in the field. We begin by deriving a precise discrete-time input-output relationship that incorporates fractional delay and Doppler, thereby addressing gaps in previous OTFS-based radar studies~\cite{zacharia2023fractional}. Building on this, we enhance the accuracy of fractional parameter estimation by employing an interpolation method based on the RRC autocorrelation function. Simulation results show that, compared to traditional linear interpolation, this approach achieves better accuracy and robustness.

The remainder of the paper is organized as follows. Section II first presents the formulation of the problem, and then introduces the generalized window function-based OTFS signal. Section III describes the proposed radar system model. 
Section IV provides the simulation results, and Section V concludes the entire paper.  
 
\setlength{\columnsep}{0.25in}
\section{Background and Proposed Model}
\subsection{Assumptions for environment and transmitted signal}

We examine coexistence challenges between radar and wireless communications. A car, equipped with an oscillator and a millimeter-wave antenna, transmits and receives data while using the same antenna to detect reflected waves (see Fig.~\ref{fig:Car_and_reflected_echos}). These reflections may come not only from the car ahead but also from ground structures or adjacent vehicles. Since signals in the same frequency band are reflected, continuous wave systems pose challenges, making pulse systems more suitable. In this scheme, the system transmits for a set time, then stops and switches to receive mode.
\begin{figure}
    \centering
    \includegraphics[width=1\linewidth]{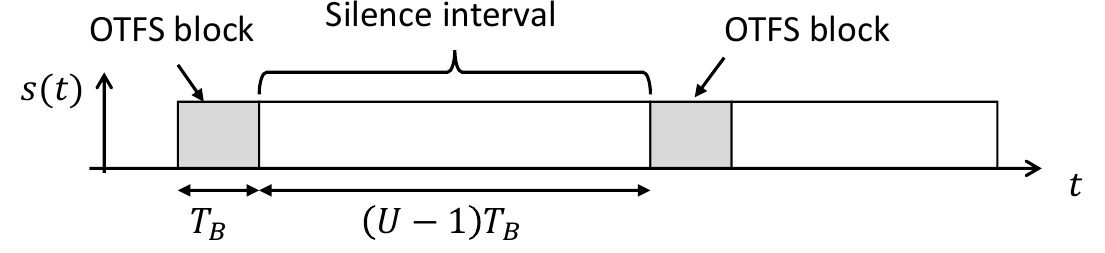}
    \caption{Structure of the Transmitted Signal}
    \label{fig:transmit_signals}
\end{figure}

The time slot of the transmitted signal is shown in Fig.~\ref{fig:transmit_signals}.
Let $T_B $ be the interval of an OTFS block, and the transmitter sends the next OTFS block after a silent interval with duration $(U-1)T_B$. Then, the pulse repetition interval (PRI) is $UT_B$.
This corresponds to the rectangular window case, and the signal duration increases if a different window is applied. 

Assume the antenna receives a total of \( P \) reflected echoes, where \( P \) is small (e.g., 1, 2, $\ldots$,  5). Let \( \alpha_i \), \( t_{D,i} \), and \( f_{D,i} \) denote the attenuation, propagation delay, and Doppler shift of the \( i \)-th reflected echo, respectively. Given the transmitted signal \( s(t) \), the received signal, in the absence of noise, is expressed as:
\footnote{
We introduced a phase shift $ - \pi f_{D,i} t_{D,i} $ so that $r(t)$ and
its Fourier transform 
$R(f) = \sum_{i=0}^{P-1} \alpha_i S(f-f_{D,i}) {\rm e}^{ -j 2\pi t_{D, i} (f - f_{D,i}/2) }$
has a symmetric relation.
}
\begin{align}
    r(t) = \sum_{i=0}^{P-1} \alpha_i s(t-t_{D,i}) {\rm e}^{j 2\pi f_{D, i} (t - t_{D,i}/2) }. \label{r(t)}
\end{align}
Assume the ranges of $t_{D, i}$ and $f_{D,i}$ are $[t_{D,\min}, t_{D,\max}]$ and $[-f_{D,\max}, f_{D,\max}]$. Because of the signal construction shown in Fig.\ref{fig:transmit_signals}, it is required that $t_{D,\min} > T_B$ and $t_{D,\max} < (U-1) T_B$. For the limitation of $f_{D,\max}$ comes from the sampling rate of the signal $T_s$, i.e.,
$2f_{D, \max} < 1/T_s$. This restriction is not a significant issue because, in real-world scenarios, the Doppler shift is much smaller than $1/T_s$.

Delay-Doppler estimation differs between radar and channel estimation for wireless communications. In communication, frame synchronization is assumed to have been established for channel estimation in OTFS. The goal is channel equalization, which occurs in the delay-Doppler domain. In this case, ‘delay’ refers to the delay spread, which is defined as the difference between the arrival times of the earliest and latest multipath components.
In contrast, the target’s presence is unknown in radar, so the goal is to detect the target and estimate its delay (corresponding to distance) and Doppler shifts (related to velocity), with high accuracy being beneficial. In our system, $t_{d,i}$ denotes the round-trip time of the reflected waves received by car A.

Unlike the overly idealized received signal models in~\cite {wu2023dft} and~\cite {zacharia2023fractional}, which are based on~\cite{Raviteja2018} and treat the delay as periodic, our model assumes a non-periodic transmitted signal. A non-periodic signal can better simulate real-world echo conditions, making the model applicable to more complex scenarios. This makes our signal model more general, with the number of received signal samples being approximately \((U-1)NM\), as shown in  Fig.~\ref{fig:transmit_signals}. Another point worth mentioning is that, although signals consisting of both pilot and data signals are more common in real-world communication, they will be discussed separately here for a clearer description of our model.
\subsection{OTFS Signal Model with Adjustable Window Function }

In OTFS systems, the data symbols are allocated on Delay-Doppler (DD) domain. Let the DD domain data symbols be $X_{\rm DD}[k,\ell]$. Then time-frequency (TF) domain symbols are defined by
\begin{align}
    X_{\rm TF}[n, m] = \frac{1}{ \sqrt{NM} } \sum_{k=0}^{N-1}\sum_{\ell=0}^{M-1} 
    X_{\rm DD}[k,\ell] W_{N}^{-n k} W_M^{m \ell},
    \label{symplecticFFT}
\end{align}
where $W_N = {\rm e}^{-j \frac{2\pi}{N} }$ is the twiddle factor of $N$-point DFT.
$ X_{\rm DD}[k,\ell] $ is periodic both in $k$ and $\ell$, i.e., 
$ X_{\rm DD}[k,\ell] =  X_{\rm DD}[k+N,\ell+M]$. 
For pilot signal $ X_{\rm DD}[0,0] = 1 $, and $ X_{\rm DD}[k,\ell] = 0 $ when $[k,\ell] \ne [0,0]$.

The representations of the transmitted signal and received samples shown below are commonly used as definitions of OTFS signals.
\begin{align}
    s(t) & = \sum_{n=0}^{N-1} \sum_{m=0}^{M-1} X_{\rm TF}[n,m] g_{\rm TX}(t-nT) \textrm{e}^{\textrm{j} \frac{2\pi}{T} m t}, \label{s(t)} \\
    Y_{\rm TF}[n,m] & = \int_{-\infty}^{\infty} r(t) 
    g^*_{\rm RX}(t-nT) \textrm{e}^{-\textrm{j} \frac{2\pi}{T} m t}
    \mathrm{d}t, \label{YTF}
\end{align}
where $g_{\rm TX}(t)$ and $g_{\rm RX}(t)$ are the transmitter's and the receiver's pulse shape; 
$T=\frac{T_B}{N}$ is the time spacing; $N$ and $M$ are the number of symbols in time and frequency domains; and $ X_{\rm TF}[m,n] $ and $ Y_{\rm TF}[m,n] $ are time-frequency (TF) domain symbols and its reconstructed ones. It is worth mentioning that while the discrete model is also a key approach, the continuous model will be used in this paper, with $x_{\rm TD}[nM+\ell]$ serving as a building block for a clearer explanation of our signal model. 

The above signal model transmits \( X_{\rm TF}[m,n] \) via the waveform \( g(t-nT) e^{j 2\pi mt/T} \), which inherently neglects the window function, rendering it unmodifiable. However, since OTFS signals are originally designed and encoded in the DD domain, we aim to design the transmitted signal directly from the DD domain as well, allowing the window function to fully realize its potential. Thus, the OTFS transmitted signal can be defined as
\begin{align}
    s(t) = \sum_{k=0}^{N-1} \sum_{\ell =0}^{M-1} X_{\rm DD}[k,\ell] h_{k,\ell}(t). 
    \label{s(t)new}
\end{align}
The \( h_{k,\ell}(t) \) in (\ref{s(t)new})  is the waveform transmitting the \( (k,\ell) \)-th symbol, which can be expressed as follows
\footnote{
The Fourier transform of $h_{k,\ell}(t)$ is given by $ \{ W(f) * X_{k,\ell}(f) \} P(f)$. 
This demonstrates that the effects of $p(t)$ and $w(t)$ are not symmetric with respect to the Fourier transform. 
Another possible choice for the waveform is $ \tilde h_{k,\ell} (t) = w(t) \{ x_{k,\ell}(t) * p(t) \}$. 
The choice between the two waveforms depends on the implementation method; 
however, the difference between the two design approaches is minimal.
},
\begin{align}
    h_{k,\ell} (t) &= \{ w(t) x_{k,\ell}(t)  \}  * p(t) , \label{hklI}
\end{align}
where $p(t)$ is the pulse shaping function, $w(t)$ is the window function, $*$ denotes the convolution operation, and $x_{k,\ell} (t) $ is the time and frequency-shifted version of a sequence of Dirac delta functions, defined by
\begin{align}
x_{k,\ell} (t) &= \sum_{n=-\infty}^{\infty} \delta \left(t -nT - \frac{\ell}{M}T \right) 
W_N^{-nk}. 
\end{align}
Since $x_{k,\ell}(t)$ is periodic with period $NT$, the Fourier transform of $x_{k,\ell}(t)$ is also a sequence of delta functions
\footnote{
The Fourier spectrum of $x_{k,\ell}(t)$ is $X_{k,\ell}(f) = 
\sum_{m = -\infty}^{\infty} \delta \left(f -\frac{m}{T} - \frac{k}{NT} \right) 
W_M^{ \ell (m + \frac{k}{N})}$. 
}. Additionally, the discrete form of \( h_{k,\ell}(t) \) can be obtained by defining\footnote{A discrete-time signal is represented using square brackets, while a continuous-time signal is represented using round brackets.
Let $\widehat W(f) = \sum_{k} W(f-\frac{k}{T_s}) = \sum_{\ell} \hat{w}[\ell] {\rm e}^{ {\rm j} 2\pi f \ell T_s}$ be the discrete-time Fourier transform (DTFT) of $\hat{w}[\ell]$.
Then we have
$w( \ell T_s) = \int_{-\infty}^\infty W(f) {\rm e}^{{\rm j} 2\pi f \ell T_s} df
= \int_{-1/(2T_s)}^{1/(2T_s)} \widehat{W}(f) {\rm e}^{{\rm j} 2\pi f \ell T_s} df
= x[\ell]/T_s
$. 
} 
$\hat{w}[\ell] = w(\ell T_s) $, as shown below,
\begin{align}
    h_{k,\ell} (t) &= \sum_{n=-\infty}^{\infty} 
    \hat{w}[nM+\ell]
    p \left(t -nT - \frac{\ell T}{M}  \right) 
    W_N^{-nk}.  \label{hklt}
\end{align}

\begin{figure}
    \centering
    \includegraphics[width=1\linewidth]{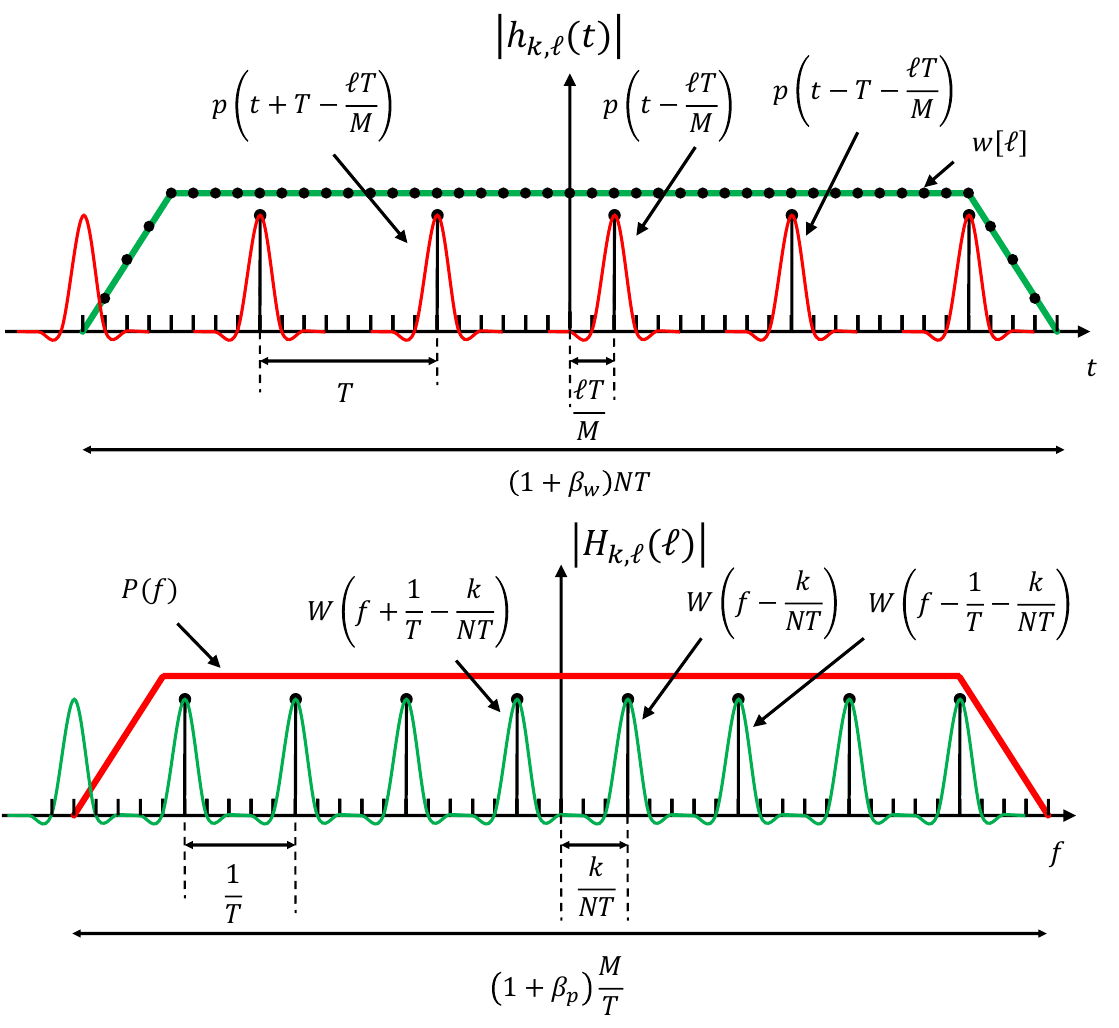}
    \caption{ Construction of $h_{k,\ell}(t)$ with the shape of $p(t)$ and $\hat{w}[\ell]$ and their spectra  ($N=5, M=8$)}
    \label{fig:h_kl-crop}
\end{figure}

To further illustrate the proposed signal model, a typical example of $h_{k,\ell}(t)$ and its Fourier transform is given in Fig.~\ref{fig:h_kl-crop}.
In this example, the window function is a rectangular window defined as $w(t) = 1$ for $|t|< \frac{T_B}{2}$ and $w(t) =0$ otherwise. Similarly, the pulse shaping filter is a rectangular pulse, where $p(t)=1$ for $0\le t < T_s$ and $p(t) = 0$ otherwise. The parameters $\beta_w$ and $\beta_p$ in this figure represent the roll-off factors of the window function and the pulse shaping filter, respectively, which are explained in Section~\ref{sect:pulse_and_window}.
By further substituting Equation  (\ref{hklt}) into (\ref{s(t)new}) and extending the range in $x_{\rm TD}$ to include all integers, the final transmitted signal model is obtained, as shown below, 
\begin{align}
   s(t) 
     &= 
  \sum_{\ell=-\infty}^{\infty} \hat{w}[\ell] x_{\rm TD}[\ell] p(t- \ell T_s) \notag \\
     &= 
    \sum_{\ell=-\infty}^{\infty} \tilde x_{\rm TD}[\ell] p(t- \ell T_s), 
    \label{s(t)new_2}
\end{align}
where $\tilde x_{\rm TD}[\ell] = \hat{w}[\ell] x_{\rm TD}[\ell]$, and $ x_{\rm TD} $ can be expressed as
\begin{align}
 x_{\rm TD} [nM+\ell] = \frac{1}{\sqrt{N}}\sum_{k=0}^{N-1} X_{\rm DD}[k,\ell] W_{N}^{-nk}.
\label{x[nM+l]}
\end{align}

It is worth noting that when the window function  $\hat{w}[\ell]$ in (\ref{s(t)new_2}) is rectangular, it appears to have the same form as (\ref{s(t)}). However, the pulse width of $\hat{w}[\ell]$ is $NT$, while the pulse width of $g_{\rm TX}$ is $T$, highlighting the difference between them.  This also indicates that the traditional OTFS model indeed neglects the window function.

A rectangular window function results in a comb-like frequency spectrum whose individual peaks take the form of sinc functions, making the estimation of Doppler frequencies more challenging. The proposed waveform, on the other hand, is used to shape the comb-like spectrum in the frequency domain, thereby facilitating the estimation of delay and Doppler parameters. Compared to the recently proposed ODDM scheme~\cite{TongODDM2024}, the proposed approach offers enhanced design flexibility by permitting the use of various window functions and pulse shaping waveforms shown as (\ref{hklt}). This flexibility allows for signal design to be optimized according to specific application requirements, thereby reducing bit error rates, mitigating inter-path interference, and improving overall system robustness.

\section{Radar sensing based on proposed model}
\subsection{ Proposed radar system model }

Building on the proposed OTFS signal model with adjustable window function, this section develops a radar model to extract Doppler and delay parameters. 
The core concept of delay-Doppler estimation involves computing the cross-ambiguity function between the transmitted and received signals:
\begin{align}
    A_{r,s}(\tau,\nu) 
    =
    \int_{-\infty}^{\infty}
    r(t + \frac\tau2) \overline s(t-\frac\tau2) e^{-j 2\pi \nu t} {\rm d} t,
\end{align}
where $\overline{z}$ denotes the complex conjugate of $z$. The peaks of the absolute value of $A_{r,s}(\tau,\nu)$ serve as candidates for the delay and Doppler estimates. The continuous-time and continuous-frequency cross-ambiguity function serves as a fundamental model but is computationally expensive. To address this limitation, we employ the discrete-time and discrete-frequency cross-ambiguity function instead.

The discrete-time version of the received signal is obtained by applying the matched filter \( \hat p(t) := \overline p(-t) \) to the output, sampled at \( t = \ell T_s \), as follows:
\begin{align}
    y_{\rm TD}[\ell ] &= \int_{-\infty}^{\infty} r(t) \hat p (t-\ell T_s) {\rm d} t, \label{yTD}
\end{align}
where $\ell $ is from $\lfloor (1+\beta_w) NM \rfloor$ to $\lfloor ( U - (1+\beta_w) ) NM \rfloor $. Then the discrete-time discrete-frequency cross-ambiguity function is defined as
\footnote{The variables $k$ and $\ell$ correspond respectively to the Doppler and the delay, while $\tau$ and $\nu$ correspond to the delay and the Doppler, respectively.} 
\begin{align}
    A_{y,x}[k,\ell] = 
    \frac1{\sqrt{NM}}
    \sum_{\ell'} y_{\rm TD}[\ell' + \ell] \tilde x_{\rm TD} [\ell'] W_{NM}^{k \ell'}. 
    \label{Ayx}
\end{align}
This is the DFT of the product $y_{\rm TD}[\ell' + \ell] \tilde x_{\rm TD} [\ell'] $ with respect to $\ell'$.
The frequency bin is $1/(NMT_s) = 1/T_B$. 

The relation between the discrete-time received signal and the transmitted signal is obtained by substituting (\ref{s(t)new_2}) into (\ref{r(t)}) and then substituting the result into (\ref{yTD}) as 
\begin{align}
    y_{\rm TD}[\ell] 
    &= \sum_{i=0}^{P-1} 
    \alpha_i \sum_{\ell'}  \tilde{x}_{\rm TD} [\ell'] e^{j \pi f_{D,i} ( \ell + \ell') T_s}  \notag\\
    & \quad \cdot A_{p,p} (( \ell-\ell') T_s - t_{D,i} , - f_{D, i} ) . 
\end{align}

We consider that $|f_{D,i}|$ is much smaller than $1/T_s$. In this case, we can approximate $ A_{p,p} (( \ell-\ell') T_s - t_{D,i} , - f_{D, i} ) \approx p * \hat{p}(( \ell-\ell') T_s - t_{D,i}) $. Let $\bm{\tilde{x}}_{\rm TD}$ and $\bm{y}_{\rm TD}$ be the column vector of $\tilde{x}_{\rm TD}[\ell]$ and $y_{\rm TD}[\ell]$. Then, we have the following lemma.

\begin{lemma}
The vector form of the samples of the received signal is given by 
\begin{align}
    \bm{y}_{\rm TD} = \sum_{i=1}^P \alpha_i H(t_{D,i}, f_{D,i}) \bm{x}_{\rm TD}, 
    \label{vec_yTD}
\end{align}
where
\footnote{
In the case of integer-valued delay and Doppler, the matrix $H$ is a product of a circular delay matrix and the diagonal matrix $D(f_D)$ with $f_D = \frac{k_D}{NT}$.
$H(t_D, f_D)$ can be expressed as a product 
${\rm Toep}'(t_D, f_D) D(f_D)$ but 
$ ( {\rm Toep}'(t_D, f_D) )_{i,j}
= {\rm e}^{ {\rm j} \pi f_D (i-j) T_s}
p* \hat p( (i-j) T_s - t_{D} )
$ includes an extra phase term.
}
\begin{align}
   H(t_{D}, f_{D}) = D({\scriptstyle \frac{f_D}2}) {\rm Toep}(t_{D}) D({\scriptstyle \frac{f_D}2}) 
   \label{H-mattrix}
\end{align}
and ${\rm Toep}(t_{D})$ is a Toeplitz matrix whose $(i,j)$ element is given by 
$p* \hat p( (i-j) T_s - t_{D} )$ and $ D(f_D) = {\rm diag} ( [ 1, {\rm e}^{{\rm j} 2\pi f_D T_s},  {\rm e}^{ {\rm j} 4 \pi f_D T_s}, \ldots, {\rm e}^{ {\rm j} (L-1) 2\pi f_D T_s} ] ) 
$. Due to space limitations, we omit the proof.
\end{lemma}

\textbf{Remark 1}
The values of \( y_{\rm TD}[\ell] \) are observed only for \(\ell = \lfloor (1+\beta_w)NM \rfloor, \dots, \lfloor (U-(1+\beta_w))NM \rfloor\), where \( \beta_w \) is the excess bandwidth ratio. Similarly, the nonzero interval of \( x_{\rm TD}[\ell] \) is limited to \( \ell = 0 \) through \( \lfloor (1+\beta_w)NM \rfloor - 1 \). Consequently, the required matrix \( H \) is wide. However, for notational simplicity, both \( y_{\rm TD}[\ell] \) and \( x_{\rm TD}[\ell] \) are treated as vectors of size \( L = UNM \). In this form, the first \(\lfloor (1+\beta_w)NM \rfloor\) elements of \( y_{\rm TD}[\ell] \) are zero but correspond to unobservable regions. This adjustment is purely for notation.

\textbf{Remark 2}
In OTFS-based radar models~\cite{wu2023dft, zacharia2023fractional}, the Toeplitz matrix in (\ref{H-mattrix}) is replaced by a circulant matrix. To validate this replacement, the transmitted signal must be cyclic. 
However, even if the transmitted signal is periodic, a Toeplitz matrix cannot be strictly replaced by a circulant matrix. This is demonstrated in the following lemma.

\begin{lemma}
Suppose that the transmitted signal $s(t)$ is a periodic signal, given by $\hat s(t) = \sum_{i} s(t-i L T_s)$. Then, the samples of the received signal is replaced by
\begin{align}
    y_{\rm TD}[\ell ] 
    &=
    \sum_{i=0}^{P-1} 
    \alpha_i \sum_{\ell'}  \tilde{x}_{\rm TD} [\ell'] {\rm e}^{j \pi f_{D,i} (\ell+\ell') T_s}\notag\\
    & 
    \cdot \sum_{q=-\infty}^\infty
    A_{p,p}((\ell-\ell'-qL)T_s -t_{d,i} , -f_{D,i}) {\rm e}^{j \pi f_{D,i} q L T_s}   
\end{align}
\end{lemma}

For clarity, let us restrict the summation over $q$ to the values $q=0$ and $q=-1$. Assume $|f_{D,i}|$ is sufficiently small. Then, the summation reduces to
\begin{align}
&   p*\hat p((\ell-\ell')T_s -t_{d,i}) \notag\\
& + p*\hat p((\ell-\ell'+L)T_s -t_{d,i} ) {\rm e}^{-j \pi f_{D,i} L T_s}.
\end{align}
The first term represents the lower triangular elements, while the second represents the upper triangular elements. Thus, the upper triangular elements exhibit a phase shift. This slight difference prevents the matrix from being circulation and, consequently, from being diagonalized by the DFT matrix.
\subsection{Pulse shapes and window functions}

\label{sect:pulse_and_window}
This subsection defines several pulse-shaping filers and window functions. Denote a rectangular function as ${\rm rect}(x) = 1$ for $|x|<1/2$, ${\rm rect}(x) = 0$ for $|x|>1/2$, and ${\rm rect}(x)=1/2$ if $x=\pm 1/2$. A sinc function is defined by ${\rm sinc}(x) = \frac{\sin \pi x}{\pi x}$ for $x\neq 0$ and ${\rm sinc}(x)=1$ for $x=0$. The Raised Cosine (RC) function with roll-off factor $\beta$ is defined by
\begin{align}
    &{\rm RC}(f ; \beta) \notag\\
    &=
    \begin{cases}
    1, & {\rm for}\, |f|<\frac{1-\beta}{2} \\
    \frac12\left[1+\cos\left(\frac{\pi}{\beta} \left[ |f| -\frac{1-\beta}{2}\right] \right) \right], & {\rm for}  \frac{1-\beta}2 \le |f| \le \frac{1+\beta}{2} \\
    0, & {\rm for}\, |f| > \frac{1+\beta}{2}
    \end{cases}
\end{align}
The Root Raised Cosine (RRC) function is defined by ${\rm RRC}(t) = \int_{-\infty}^{\infty} \sqrt{ {\rm RC}(f) } {\rm e}^{ {\rm j} 2 \pi t f}  {\rm d} f$.
\subsubsection{Pulse shaping filters}

We examine three types of pulse-shaping filters. The rectangular pulse is defined by $p(t) = {\rm rect}(\frac{t}{T_s})$, The sinc pulse is defined by $p(t) = {\rm sinc}(\frac{t}{T_s} )$, and the rrc pulse with a roll-off factor $\beta_p$ is defined by $p(t)  = {\rm RRC}( \frac{t}{T_s} ; \beta_p )$.
\subsubsection{Window functions}

We examine two types of window functions. The rectangular window is given by \( w(t) = {\rm rect}\left(\frac{t}{NT}\right) \). The rrc window is expressed as \( w(t) = {\rm rrc}\left(\frac{t}{(1+\beta_w)NT}; \beta_w \right) \).
To the best of the authors' knowledge, the rrc window function has not been previously examined, and we will show that this choice is promising.
\subsubsection{Orthogonality condition}

(\ref{s(t)new}) can be viewed as an expansion of $s(t)$ using base functions $h_{k,\ell}(t)$. The orthogonality property is of crucial importance for the base functions. We have the following theorem.
\begin{theorem} 
\label{theorem_orthogonality}
If the pulse shaping filter $p(t)$ satisfies Nyquist condition after the matched filtering, i.e.,
\begin{align}
p * {\hat p(t) } \Big|_ {t= \ell  T_s} 
= \begin{cases}
    1, & \ell = 0, \\
    0, & \ell \neq 0,
\end{cases}
    \label{NyquistCondition-p}
\end{align}
and if the Fourier transform of the window function $w(t)$ satisfies the Nyquist condition in the frequency domain after the matched filtering in the frequency domain, i.e., 
\begin{align}
\left. W* {\hat W(f)} \right|_{ f = \frac{k}{T_B}} = \begin{cases}
    1, & k = 0, \\
    0, & k \neq 0,
\end{cases}
    \label{NyquistCondition-w}
\end{align}
where $\hat W(f) = \overline{W}(-f)$, then $h_{k,\ell}(t)$ are orthonormal, i.e.,
    \begin{align}
        \int_{-\infty}^{\infty} h_{k,\ell}(t) h_{k',\ell'}^*(t) {\rm d} t 
        = \delta_{k,k'} \delta_{\ell, \ell'}.
    \end{align}
\end{theorem}
Due to space constraints, the proof is omitted.

\begin{figure}
    \centering
    \includegraphics[width=0.9\linewidth]{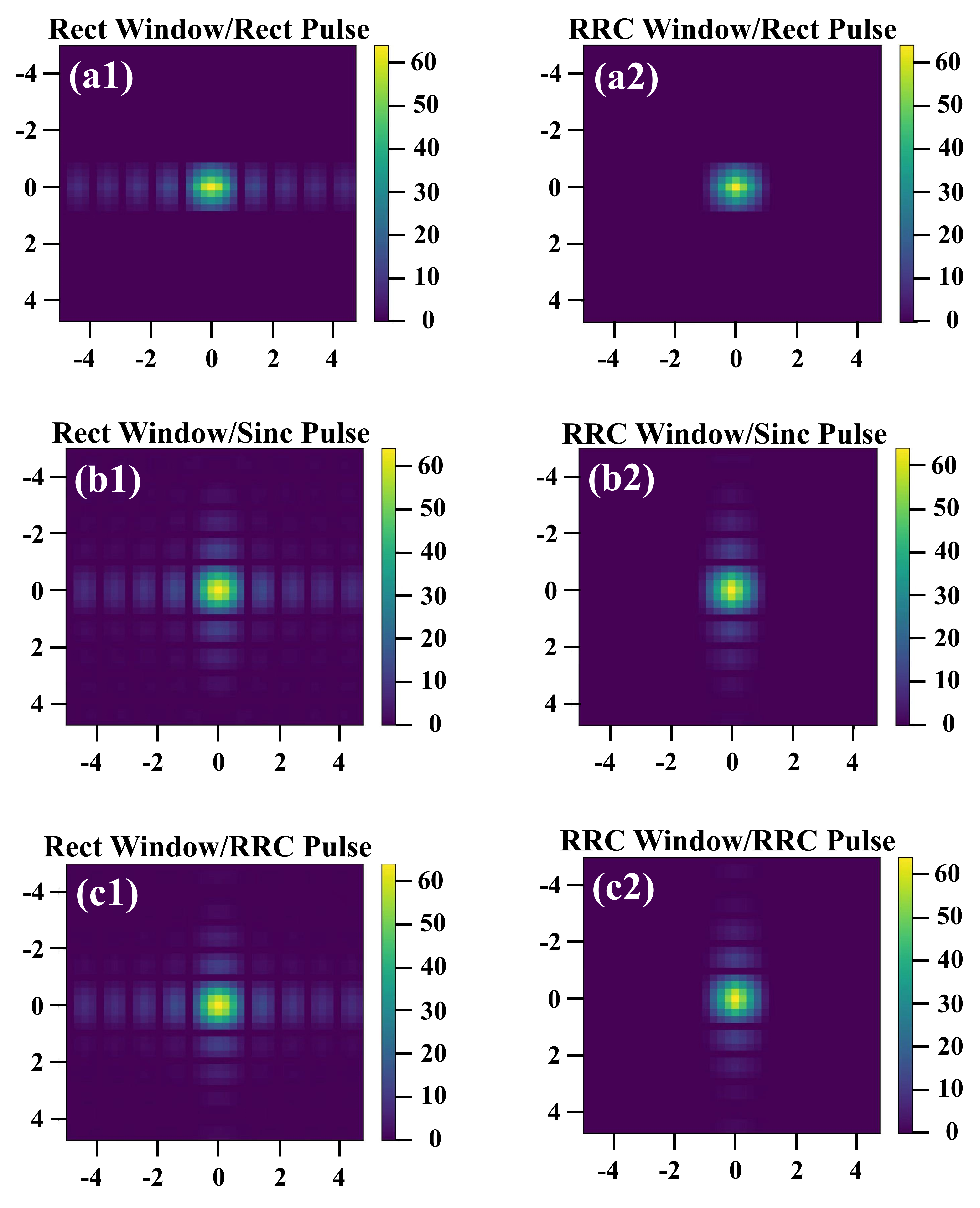}
    \caption{The fine ambiguity function $A_{s,s}(\tau, \nu)$ for 
     $-5T_s<\tau<5T_s$ and $-5/T_B < \nu < 5/T_B$.
     (a) Rectangular pulse, (b) sinc pulse and (c) rrc pulse. The roll off factor is 25\%.
     (x1) denotes a rectangular window, while (x2) means an rrc window. }
    \label{fig:ambiguity-new}
\end{figure}

The ambiguity functions of $h_{0,0}(t)$ with different pulses and window functions are shown in Fig.~\ref{fig:ambiguity-new}. The horizontal axis shows the Doppler and vertical axis shows the delay.
When using a rectangular window, as shown in Fig.~\ref{fig:ambiguity-new} (a1), (b1), and (c1), the ambiguity function of any pulse exhibits small oscillations in the Doppler direction. In contrast, when using a rectangular pulse with the proposed RRC window, as shown in Fig.~\ref{fig:ambiguity-new} (a2), no oscillations of the ambiguity function are observed. This is a crucial characteristic that simplifies signal processing in the DD domain. 

\section{Estimation Algorithm and Number simulation}
\subsection{Estimation Algorithm}

The delay and Doppler frequency to be estimated are defined as $t_{d,p} = (l + \varepsilon_t) T_s$ and $f_{d,p} = \frac{k + \varepsilon_f}{N T}$, where $l$ and $k$ are the integer parts, $\varepsilon_t$ and $\varepsilon_f$ are the fractional parts, and $T = M T_s$. Accurate delay and Doppler estimation is usually performed in two steps - identifying candidate $[\hat{k}, \hat{\ell}]$ and then focusing on $[\hat{\varepsilon_f},\hat{\varepsilon_t}]$ around the candidate $[\hat{k}, \hat{\ell}]$ using the cross-ambiguity function.
For the first step, as Algorithm~1 shows, we aim to obtain the matrix
$\begin{bmatrix}
| A_{y,x}[k,l] | & | A_{y,x}[k,l+1] |\\
| A_{y,x}[k+1,l] |& |A_{y,x}[k+1,l+1] |
\end{bmatrix}$
,which can be denoted as
$\begin{bmatrix}
A_{00} & A_{01} \\
A_{10} & A_{11}
\end{bmatrix}$
. The step has been extensively studied in many works~\cite{wu2023dft}. Thus, we skip this step and directly use the precise 2×2 matrices as input in the subsequent estimation process. 

For the second step, the total error function is first defined as
\begin{equation}
L(\alpha, \varepsilon_t, \varepsilon_f) = \sum_{i=0}^{1} \sum_{j=0}^{1} \left| A_{ij} - \alpha \cdot A_{ss}(i - \varepsilon_t, j - \varepsilon_f) \right|^2.
\label{22}
\end{equation}
The $\alpha$ is the attenuation coefficient, and the $A_{ss}$ is the central region ambiguity function of the pilot signal, well approximated by
\begin{equation}
{A}_{ss}(\tau, v) \approx p \ast {\hat p(\tau)} \cdot W \ast {\hat W(\nu)},\quad |\tau| \leq T_s, \, |\nu| \leq \frac{1}{N T}. 
\label{23}
\end{equation}
Our objective is to obtain the $(\hat{\alpha}, \hat{\varepsilon}_t, \hat{\varepsilon}_f)$ meeting $\displaystyle \arg\min_{\alpha, \varepsilon_t, \varepsilon_f} L(\alpha, \varepsilon_t, \varepsilon_f)$. For traditional linear interpolation, the $W \ast {\hat W(\nu)}$ can be expressed as
\begin{equation}
W * {\hat W(\nu)} =
\begin{cases}
1 - |\nu|N T, & |\nu| \leq \dfrac{1}{N T}, \\
0, & \text{otherwise}.
\end{cases}
\label{24}
\end{equation}
However, since the proposed RRC window function exhibits an approximately triangular but non-linear shape, conventional linear interpolation may introduce estimation errors. To enhance accuracy, we propose an interpolation method based on the autocorrelation function of the RRC window, explicitly given by

{\small
\begin{equation}
\label{eq:window_function}
W * {\hat W(\nu)} =
\begin{cases}
\begin{aligned}
& \frac{1}{2} \cos(b \nu)(1 - 2\nu - 2a) + \frac{1}{2b} \sin(b - b \nu - 2 b a) \\
& \quad - \frac{3}{2b} \sin(-b\nu) + 2a - \nu,\quad \text{for: } 0 \leq \nu \leq -a + \frac{1}{2}\\
\end{aligned} \\[1.5ex]
\begin{aligned}
& \frac{2}{b} \sin\left(\frac{b}{2} - b a\right) + 2a - \nu,\quad \text{for: } -a + \frac{1}{2} < \nu \leq 2a\\
\end{aligned} \\[1.5ex]
\begin{aligned}
& \frac{2}{b} \left[ \sin\left(\frac{b}{2} - b a\right) - \sin(-2 b a + b \nu) \right] \\
& + \frac{1}{4b} \left[ \sin(-2 b a + b \nu) - \sin(2 b (a - \nu) + b \nu) \right] \\
& + \frac{1}{2} \cos(b \nu - 2 b a)(-2a + \nu),\quad \text{for: } 2a < \nu \leq a + \frac{1}{2}\\
\end{aligned} \\[1.5ex]
\begin{aligned}
& \frac{1}{2} \left[ \frac{1}{b} \sin(b - b \nu) + \cos(-2 b a + b \nu)(-\nu + 1) \right], \\
&\quad \text{for: } a + \frac{1}{2} < \nu \leq 1
\end{aligned} \\[1.5ex]
0, \quad \text{for: otherwise, }
\end{cases}
\end{equation}
}where $a = \frac{1 - \beta}{2(1 + \beta)}$, $b = \frac{\pi (1 + \beta)}{2\beta}$, and $\beta$ is chosen to be 0.25. This method can describe the signal structure more precisely and thus improves the estimation performance.
\begin{algorithm}
    \caption{Coarse Delay and Doppler Estimation}
    \begin{algorithmic}[1]
    \REQUIRE $x_\mathrm{TD}[\ell]$ for the signal transmission duration and $y_\mathrm{TD}$ for the reception period. Set the parameter $\mathrm{MAX\_PATHS}$.
    \STATE Compute the ambiguity function (\ref{Ayx}) using the $NM$-point FFT algorithm for each  
    $y_\mathrm{TD}[\cdot + \ell] x_\mathrm{TD}[\cdot]$.
    \STATE Select the $[k,\ell]$ pairs corresponding to the $\mathrm{MAX\_PATHS}$ largest absolute values of $ A_{y,x}[k,\ell] $. This set, $\{ [k,\ell] \}$, is referred to as the candidate list.
    \STATE For each candidate $[k,\ell]$,  
    check whether $ | A_{y,x}[k,\ell] |$ is the maximum value within a $5\times 5$ neighborhood centered at $[k,\ell]$.  
    If not, remove $[k,\ell]$ from the candidate list.
    \STATE (2D-CA-CFAR)  
    For each candidate $[k,\ell]$,  
    check whether $ | A_{y,x}[k,\ell] |$ exceeds the average of $ | A_{y,x}[k \pm i,\ell\pm j] |$, where  
    $(i,j)$ takes values in $\{-2,-1,0,1,2\} \times \{-2,-1,0,1,2\} \setminus (0,0)$.  
    If not, remove $[k,\ell]$ from the candidate list.
    \RETURN The candidate list.
    \end{algorithmic}
\end{algorithm}
\begin{algorithm}
    \caption{Fractional Delay and Doppler Estimation}
    \begin{algorithmic}[1]
    \STATE The precise 2×2 matrices of $[k,\ell]$ , corresponding to the integer components of delay and Doppler, are used as input.
    \STATE Based on interpolation function, the optimize.minimize is used to minimize the total squared error between the interpolation result and input 2×2 matrix, estimating the fractional delay/Doppler and attenuation coefficients. 
    \STATE \textbf{return} The RMSE of the estimated result and true values.
    \end{algorithmic}
\end{algorithm}

\subsection{Numerical simulation}

In this simulation, the DD domain symbols are assumed to be OTFS pilot signals, and the system parameters are set to \( N = M = 8 \).  
The RRC window function and rectangular pulse waveform are adopted due to their minimal ambiguity function oscillations, as shown in Fig.~\ref{fig:ambiguity-new}.  
The number of unknown propagation paths ranges from 1 to 5, and each scenario is averaged over 100 independent runs.  
The 2×2 grid centered at each ambiguity peak is assumed to be known and is utilized by Algorithm~2 for refined parameter estimation.  

The goal is to compare the estimation performance of the traditional linear interpolation method and the proposed method based on the RRC autocorrelation function, in terms of attenuation coefficients (\(\alpha\)), fractional delays (\(\varepsilon_t\)), and fractional Doppler shifts (\(\varepsilon_f\)).  
The accuracy of the estimate is quantified by the root mean square error (RMSE), defined as
\[
\text{RMSE} = \sqrt{ \frac{1}{N_\mathrm{sim} \cdot P} \sum_{n=1}^{N_\mathrm{sim}} \sum_{p=1}^{P} \left( \hat{x}_{n,p} - x_{n,p} \right)^2 },
\]where \( N_\mathrm{sim} \) is the number of Monte Carlo iterations and \( P \) is the number of paths.
The symbol $x$ represents $\alpha, \varepsilon_t$ or $\varepsilon_f$, and $\hat x$ represents its estimated value.

\begin{figure}
    \centering
    \includegraphics[width=1\linewidth]{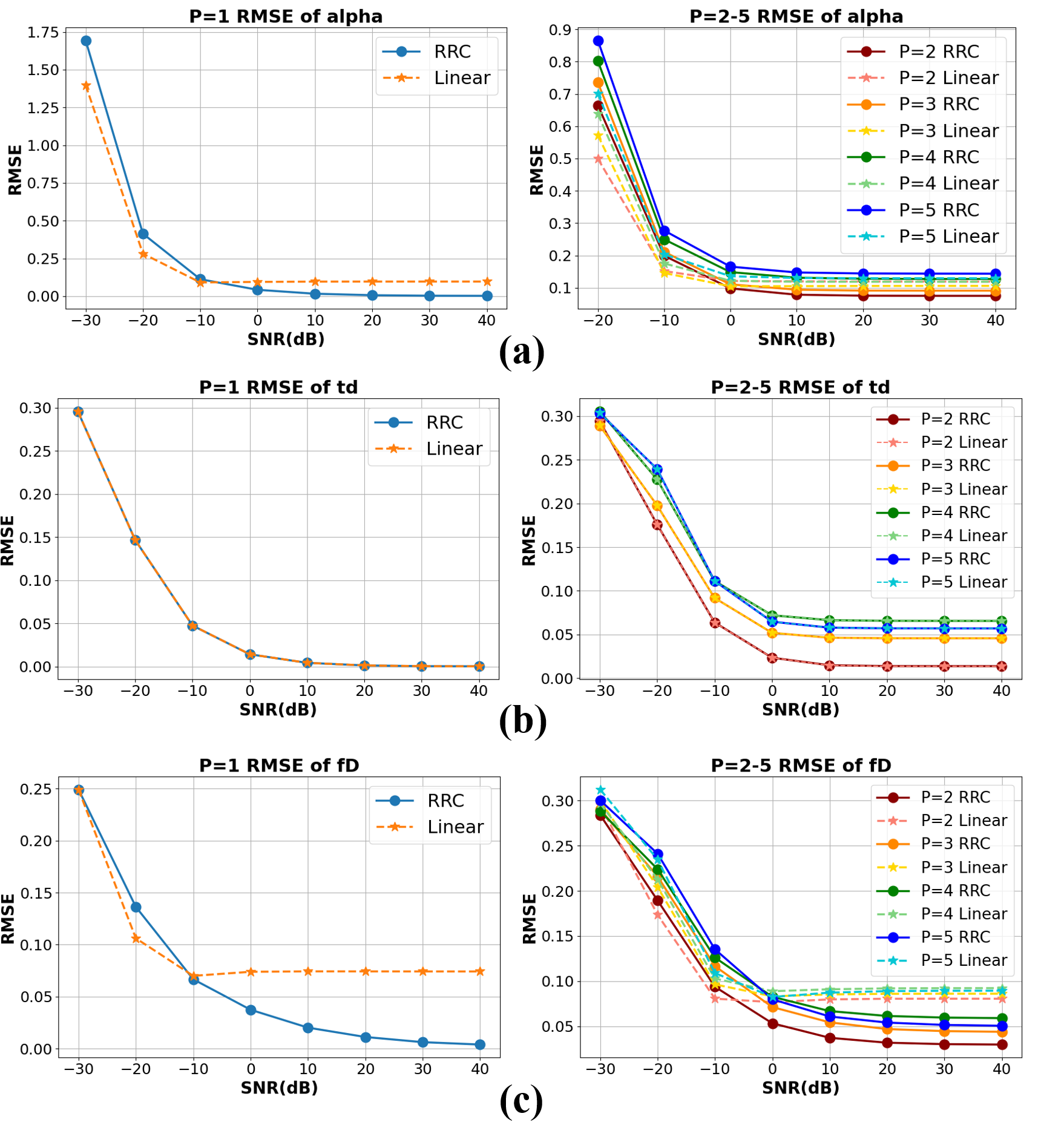}
    \vspace{-10pt}
    \caption{Comparison of estimation accuracy between linear interpolation and the proposed RRC autocorrelation-based interpolation: (a) attenuation coefficients, (b) fractional delays, (c) fractional Doppler shifts.}
    \label{fig.5}
    \vspace{-10pt}
\end{figure}

\begin{figure}
    \centering
    \includegraphics[width=1\linewidth]{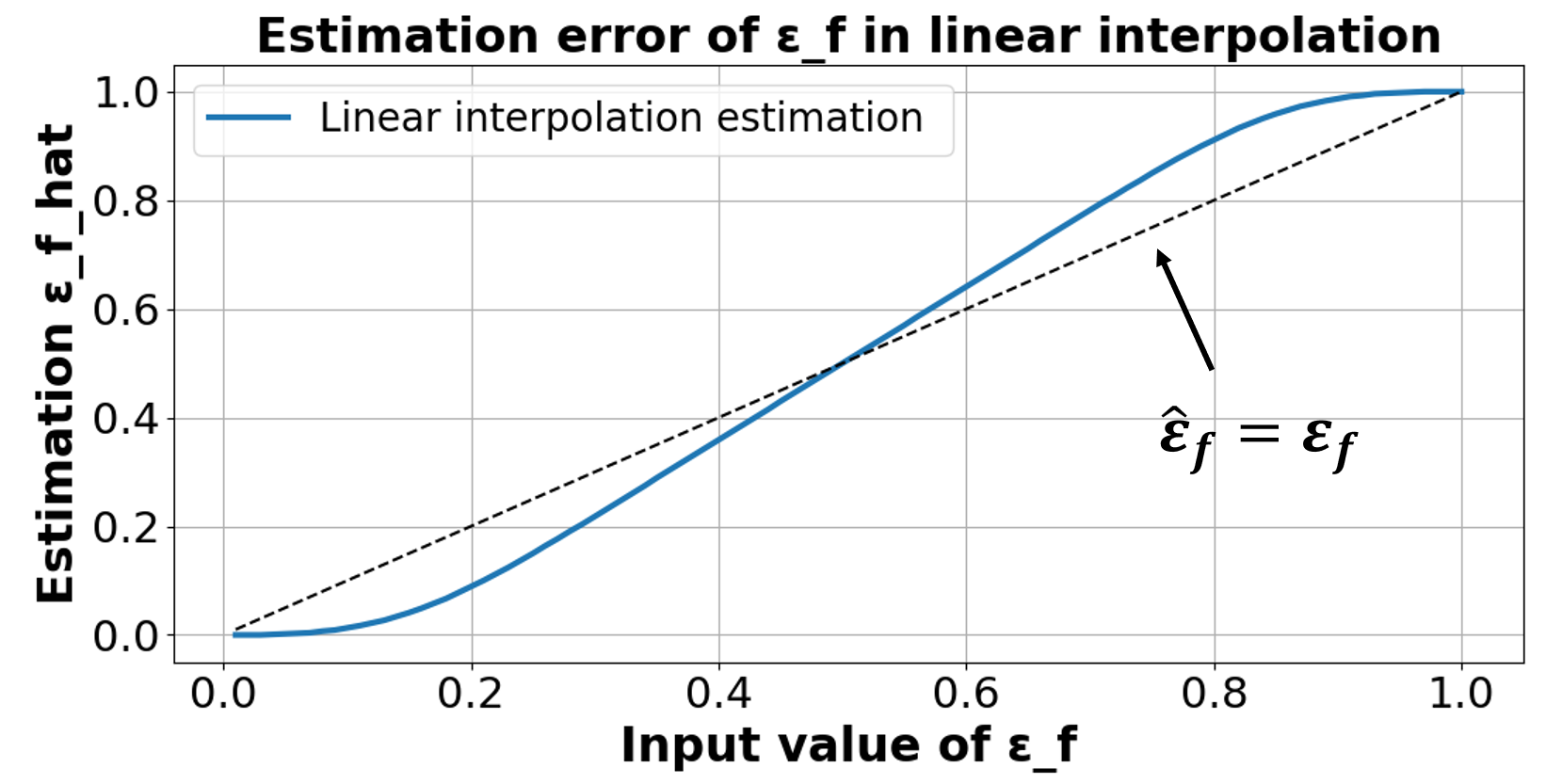}
    \vspace{-5pt}
    \caption{Estimation error analysis of fractional Doppler shifts using linear interpolation for OTFS signals with the RRC window function in the noise-free case. The input $ \varepsilon_f $ ranges from 0.01 to 0.99 with a step size of 0.01.}
    \label{fig.6}
    \vspace{-13pt}
\end{figure}

Fig.~\ref{fig.5} shows that for the estimation of $\alpha$\ and $\varepsilon_t$, the proposed method performs comparably to linear interpolation. However, when estimating $\varepsilon_f$, the proposed method demonstrates noticeably better performance. This improvement can be attributed to the fact that the RRC window function is a nonlinear function involving trigonometric components, which linear interpolation cannot accurately approximate, thereby introducing non-negligible errors. This observation is further supported by the noise-free simulation with linear interpolation and $\varepsilon_t = 0$, as shown in Fig.~\ref{fig.6}.

\section{Conclusion}

In this work, we propose a pulse radar system based on OTFS signals, which enables flexible selection of window functions and waveform shaping. We introduce the RRC window due to its low ambiguity-function oscillations when combined with a rectangular pulse. Based on this waveform design, we propose an RRC autocorrelation-based interpolation method, which significantly outperforms conventional linear interpolation in fractional Doppler estimation.
\bibliographystyle{IEEEtran}
\bibliography{new_input.bib}

\end{document}